# Plasma lenses for ultrashort multi-petawatt laser pulses

J.P. Palastro, D. Gordon, B. Hafizi, L. A. Johnson, J. Peñano, R. F. Hubbard, M. Helle, and D. Kaganovich

*Naval Research Laboratory, Washington DC 20375-5346, USA*

## Abstract

An ideal plasma lens can provide the focusing power of a small f-number, solid-state focusing optic at a fraction of the diameter. An ideal plasma lens, however, relies on a steady-state, linear laser pulse-plasma interaction. Ultrashort multi-petawatt (MPW) pulses possess broad bandwidths and extreme intensities, and, as a result, their interaction with the plasma lens is neither steady state nor linear. Here we examine nonlinear and time-dependent modifications to plasma lens focusing, and show that these result in chromatic and phase aberrations and amplitude distortion. We find that a plasma lens can provide enhanced focusing for 30 fs pulses with peak power up to ~1 PW. The performance degrades through the MPW regime, until finally a focusing penalty is incurred at ~10 PW.

# I. Introduction

A laser pulse propagating in plasma acquires a phase proportional to the plasma density. Plasma with density variation imparts a spatially varying phase, causing the pulse to refract. Thus with appropriate spatial structuring the plasma can, in principle, be made to mimic any linear, solid-state optical element. Plasma-based optical elements, being already ionized, have the advantage of higher damage thresholds, allowing their use at higher intensities than solid-state elements. Furthermore, plasma optics can be inexpensively and rapidly replaced, for instance, at the rep-rate of a gas jet or capillary [1,2], or flow rate of a water jet [3].

Plasma lenses, in particular, can provide the focusing power of small f-number solid-state parabolas at a fraction of the diameter [4-6]. These lenses can either be pre-structured with a density profile increasing with radius [5,6], or take advantage of nonlinear focusing within the plasma [4]. In the prior case, the lenses resemble axially truncated plasma waveguides [7]. Plasma waveguides enable laser propagation over distances unrestrained by vacuum diffraction and ionization refraction in applications including optically driven particle accelerators and radiation sources [8-13]. This resemblance facilitates techniques for both forming plasma lenses and modeling of their linear focusing properties. For instance, plasma lenses can be preformed through the gas ionization, plasma heating, and hydrodynamic expansion driven by a ~100ps Nd:YAG pulse focused onto a gas jet [1], the front edge ablation of a thin plastic capillary by a ~1 ns Nd: YAG pulse [6], or the ionization of neutral gas density depressions formed by colliding gas jet flows [14].

Here we examine plasma lenses for ultrashort pulse multi-petawatt (MPW) laser systems. In designing a MPW laser system, one confronts the practical issue that the final focusing optic is likely to be a large, one-of-a-kind parabolic mirror that fixes the f-number for all experiments [15]. Ideal plasma lenses [5] could provide flexibility in the focal geometry, operating as a concave (convex) lens to decrease (increase) the effective f-number. Schematics of focusing systems combining a parabolic mirror and plasma lens are shown in Fig. 1. The plasma density profiles for the focusing and defocusing plasma lenses are designed to refract rays inward and outward respectively. These ideal plasma lenses, however, rely on a steady-state, linear laser pulse-plasma interaction.

Ultrashort MPW pulses possess broad bandwidths and extreme intensities, and, as a result, their interaction with plasma lenses is neither steady state nor linear. We examine nonlinear and time-dependent modifications to the plasma lens focusing of ultrashort MPW pulses. We show that these modifications result in chromatic and phase aberrations and amplitude distortion. Specifically, we find that two effects can inhibit plasma lenses: First, the broad bandwidth of the pulses coupled with the lens dispersion lead to asynchronous focusing. Second, the extreme intensities result in phase aberrations and amplitude distortions induced by the nonlinear current driven in the plasma lens. We demonstrate that plasma lenses are generally susceptible to phase aberrations, while the focusing and defocusing plasma lenses are susceptible to chromatic aberrations and amplitude distortion, respectively.

The remainder of this manuscript is organized as follows. In section II we review the linear, steady-state plasma lens theory. Section III describes the time-dependent and nonlinear effects that inhibit plasma lens performance for ultrashort MPW pulses. Section IV presents ponderomotive guiding center (PGC) simulation results for two cases: a focusing plasma lens for reducing the effective f-number of a parabola, and a defocusing plasma lens for increasing the effective f-number. Section V concludes the manuscript.

**II. Linear, Steady-state Plasma Lens Theory**

We begin by reviewing the linear, time-independent properties of plasma lens focusing. Here and throughout linear refers to the laser pulse's peak normalized vector potential $a = eA_\perp / m_e c^2$ satisfying $a \ll 1$, where $e$ the elementary charge, $m_e$ the electron mass, and $c$ the speed of light. Time-independence requires two conditions to be met: linearity, so that the charge density remains constant throughout the pulse, and weak frequency dispersion in the lens. As we will show later, the latter condition can be expressed as $(c\tau/w_l)f_\# \gg 1$, where $\tau$ is the temporal full-width at half-maximum (FWHM) of the pulse, $w_l$ is the spot size at the plasma lens, and $f_\#$ is the focusing plasma lens f-number.

Unless otherwise stated, we consider a focusing plasma lens with radially parabolic electron density profiles of the form $n_{e0} = n_0 + \tfrac{1}{2}n_0''r^2$. This profile supports the stationary propagation of a transverse Gaussian electromagnetic mode of $\exp(-1)$ width $w_m = (2/\pi r_e n_0'')^{1/4}$ where $r_e$ is the classical electron radius [16]. The lens exhibits two qualitative behaviors depending on its thickness.

A thick plasma lens uses mode conversion in the plasma waveguide to exploit an otherwise non-ideal property: a laser pulse coupled into a plasma waveguide with $w_l > w_m$ will undergo spot size oscillations. During these oscillations, the spot size can reach a minimum $w_{min} \sim w_l(w_m/w_l)^2$ after a distance $z \sim \pi k_0 w_m^2/4$ and every $\pi k_0 w_m^2/2$ thereafter. The quadratic scaling of $w_{min}$ with $w_m$ allows, at least in principle, a sizable decrease in the spot size and concomitant increase in intensity. In particular, after half a spot size oscillation $f_\# = (1/4)(w_m/w_l)k_0 w_m$. As an example, we take parameters typical of Ti-Sapphire laser system $\lambda = 2\pi/k_0 = 800$ nm, $w_m = 15$ $\mu$m, $w_l = 250$ $\mu$m, and find $f_\# = 1.8$.

A short plasma lens, $\Delta < k_0 w_m^2/2$ where $\Delta$ is the lens thickness and $k_0$ the laser pulse wavenumber, imparts a quadratic phase analogous to the phase applied by a thin lens. Specifically, $\phi = -i(2\Delta/k_0 w_m^4)r^2$, giving an effective $f_\# = (1/8)(w_m^2/\Delta w_l)(k_0 w_m)^2$ provided $f_\# > (\Delta/2w_l)$. For the same parameters as above and $\Delta = 0.5$ mm, we find $f_\# = 3.1$. A short defocusing plasma lens can be considered by simply changing the sign of the electron density's radial dependence: $n_{e0} = n_0 - \tfrac{1}{2}n_0''r^2$ with the condition $r < 2n_0/n''$ implied throughout. In the remainder, we will focus on short plasma lenses.

The shorter interaction length results in less nonlinear aberration compared to thick plasma lenses.

### III. Nonlinear and Time-dependent Considerations

For MPW systems, the plasma lens diameter required for the linear focusing properties described above can be impractical. To illustrate this, we consider a parabolic mirror focused Gaussian beam incident on a plasma lens. Conservation of pulse power implies $a_f w_f = a_l w_l$, where the subscripts '$f$' and '$l$' refer to the parabola's vacuum focus and axial location of the plasma lens respectively. As the plasma lens is moved closer to the parabola, the incident intensity drops but the spot size increases, necessitating a larger plasma lens. In particular, the plasma lens diameter scales as $D_l \sim 1.9(P/P_*)^{1/2} a_l^{-1} \lambda$, where $P_* = m_e c^3 / r_e = 8.7$ GW. We have introduced a factor of 3/2 in the scaling to ensure the lens encapsulates 99% of the pulse power ($D_l \sim 3 w_l$). Expressed in this way, $D_l$ is independent of the f-number, depending only on the laser power and the acceptable vector potential, as a measure of nonlinearity, incident on the plasma lens. As an example, we take $P = 10$ PW, $\lambda = 800$ nm, and place the plasma lens at an axial position where the field is weakly relativistic, e.g. $a_l = 0.25$. The resulting plasma lens diameter is a rather large 0.65 cm.

Even if a large plasma lens could be created, one must contend with the large plasma densities and chromatic aberration. For a focusing plasma lens, the density reaches quarter critical at a radius of $r_{qc} \sim 0.82 (\Delta D_l f_\#)^{1/2}$. When $r_{qc} < D_l / 2$, the absolute Raman instability at the lens surface can limit the pulse power refracted by the lens [17-19]. This places a condition on the "usable" lens diameter, $D_l \leq 2.6 \Delta f_\#$. If the absolute Raman instability has a small gain or can be mitigated, reflection from critical density plasma limits the useable lens diameter instead, $D_l \leq 10 \Delta f_\#$.

Chromatic aberration results from the dispersion in the plasma lens coupled with the large bandwidth of an ultrashort pulse, and manifests as an asynchronous focusing.

As we will show later, asynchronous focusing leads to a significant drop in the focal intensity when $(c\tau/w_l)f_\# < 1$. Using the above expression for $D_l$, we can rewrite this condition as $1.6 a_l c\tau f_\# \lambda^{-1}(P/P_*)^{-1/2} < 1$ to illustrate two points. First, moving the plasma lens closer to the parabola to reduce $a_l$ exacerbates the effect of asynchronous focusing. Second, high power ultrashort pulses have an increased susceptibility to asynchronous focusing.

While there is a trade-off between reducing the effects of nonlinearity and asynchronous focusing, a reasonably sized plasma lens ($D_l$ < 1 mm) will necessarily be placed in a region of high field intensity ($a_l$ >1). At these intensities, the pulse acquires dynamic phase aberrations from the nonlinear electron current it drives in the plasma lens. These aberrations can significantly degrade the plasma lens focusing. Furthermore, smaller diameter plasma lenses still result in asynchronous focusing especially those with small f-numbers. As a result, neither the simple steady-state linear f-number estimates provided above, nor weakly relativistic expansions suffice [4,5]. Said differently, a plasma lens model for MPW systems must capture fully nonlinear, time-dependent modifications to the plasma density and high order optical aberrations. In the next section we will present simulations that capture these modifications. Before doing so, we examine limiting cases to obtain insight into the effects of nonlinearity and asynchronous focusing.

We express the laser pulse's normalized transverse vector potential, $\mathbf{a}_\perp$, as a carrier wave modulating a slowly varying envelope, $\hat{\mathbf{a}}$: $\mathbf{a}_\perp(\mathbf{r},z,\xi) = \frac{1}{2}\hat{\mathbf{a}}(\mathbf{r},z,\xi)e^{-ik_0\xi} + \text{c.c.}$ where $\xi = ct - z$ measures distance in the frame moving with the pulse at velocity $c$. In the modified-paraxial approximation [12,20], the envelope evolves according to

$$\left[2\frac{\partial}{\partial z}\left(ik_0 - \frac{\partial}{\partial \xi}\right) + \nabla_\perp^2\right]\hat{\mathbf{a}} = k_p^2 \rho \hat{\mathbf{a}} \quad (1)$$

where $k_p = \omega_p/c$, $\omega_p = (e^2 n_N / \varepsilon_0 m_e)^{1/2}$, $n_N$ is the density normalization, $\rho(\mathbf{r}, z, \xi) = n_e / \gamma n_N$, $n_e$ is the electron density, $\gamma = (1 + u^2)^{1/2}$, $\mathbf{u} = \gamma \mathbf{v}/c$, and $\mathbf{v}$ is the cold fluid electron velocity [21]. Chromatic aberrations arise from the combined effect of the mixed derivative, $\partial^2_{\xi z}$, on the left of Eq. (1) and $\rho$ on the right. The implicit intensity dependence of the current on the right of Eq. (1), $\mathbf{J}_\perp = -ecn_N \rho \mathbf{a}_\perp$, drives the nonlinear phase aberrations.

### A. Asynchronous Focusing

To determine the effect of asynchronous focusing, we consider Eq. (1) in the thin lens, linear regime. The thin lens approximation amounts to replacing $\rho(\mathbf{r}, z, \xi)$ by $\Delta \delta(z - z_l) \rho(\mathbf{r}, \xi)$ in Eq. (1), condensing the phase imparted to the pulse over a distance $\Delta$ into a single "phase screen" located at the center of the plasma lens, $z = z_l$. The conditions for which the thin-lens approximation holds are described in Appendix A. In this limit, one can Fourier transform Eq. (1) with respect to $\xi$ and obtain a modified-Fresnel integral for the solution of $\tilde{\mathbf{a}}(\mathbf{r}, z, \delta \omega)$ anywhere along the propagation path. Here '~' denotes a Fourier transform with respect to $\xi$ with conjugate variable $\delta \omega$. Using $n_e = n_0 \pm \frac{1}{2} n_0'' r^2$, and propagating the pulse a distance $L$, we have up to a constant phase

$$\tilde{\mathbf{a}}(r, L + z_l, \delta\omega) = \frac{k_\omega}{2\pi i L} \int \exp\left[i\frac{k_\omega}{2L}(\mathbf{r}-\mathbf{r}')^2 \mp i\frac{k_0^2 r'^2}{2k_\omega f_{pl}}\right] \tilde{\mathbf{a}}(r', z_l, \delta\omega) d^2 r' \quad (2)$$

where we have defined $k_\omega = k_0 + \delta\omega/c$ and $f_{pl} = \pm k_0^2 w_m^4 / 4\Delta$.

Both the diffractive phase and thin lens phase in Eq. (2) contribute to chromatic aberration. For a defocusing plasma lens, the frequency-dependence of the phases partially cancels, reducing the aberration. For a focusing plasma lens, on the other hand, the frequency-dependent phases add essentially doubling the aberration. In terms of the pulse profile, the increase in electron density with radius results in the local group velocity ($v_g/c = 1 - k_p^2 n_e(r) / 2 k_\omega^2 n_N$) decreasing with radius. The slippage of the pulses's radial edges with respect to its center creates a boomerang-like spatio-temporal intensity profile as the pulse approaches focus. An example is shown in Fig. (2a). This profile is in stark contrast with an ideal Gaussian profile. We refer to this effect as asynchronous focusing.

To evaluate the integral in (2) we consider a Gaussian pulse in both space and time focused by a parabola, such that up to a constant phase

$$\tilde{\mathbf{a}}(r,z_l,\delta\omega) = \sigma\pi^{1/2}\mathbf{a}_l \exp\left[-\frac{r^2}{w_l^2} + i\frac{k_0 r^2}{2R_l} - \frac{1}{4}\sigma^2\delta\omega^2\right] \quad (3)$$

where $w_l = w_f(1+z_l^2/Z_R^2)$, $R_l = z_l(1+Z_R^2/z_l^2)$, $Z_R = k_0 w_f^2/2$, and $\sigma = \tau/[2\ln(2)]^{1/2}$. Upon Taylor expanding $k_\omega^{-1} \cong k_0 - \delta\omega/c$ in the plasma lens term of Eq. (2), inverse Fourier transforming, and forming the magnitude squared, we have:

$$|\hat{\mathbf{a}}(0,L+z_{pl},\xi)|^2 = \pi\left(\frac{Z_R a_l S}{L}\right)^2 \exp\left[-\frac{2\xi^2}{c^2\sigma^2}\right]\text{erfcx}\left\{S\left[1-\frac{\xi}{Sc\sigma}-i\frac{Z_R}{L}\left(1-\frac{L}{f_{pl}}+\frac{L}{R_l}\right)\right]\right\}\bigg|^2 \quad (4)$$

where erfcx is the scaled complimentary error function and $S = c\sigma w_l^{-2} L f_{pl}/(L+f_{pl})$. We note that calculations were performed maintaining the second order term in the Taylor expansion of $k_\omega^{-1}$, but the results were not appreciably different for the parameters considered here.

The size of $S$ indicates the importance of asynchronous focusing and provides the condition $(c\tau/w_l)f_\# < 1$, which we discussed earlier. In the limit $S \to \infty$, we recover the ideal Gaussian optics result from Eq. (4):

$$|\hat{\mathbf{a}}(0,L+z_{pl},\xi)|^2 = a_l^2 \exp\left[-\frac{2\xi^2}{c^2\sigma^2}\right]\left[\left(\frac{L}{Z_R}\right)^2 + \left(1-\frac{L}{f_{pl}}+\frac{L}{R_l}\right)^2\right]^{-1}. \quad (5)$$

Figure (2b) displays a comparison of Eqs. (4) and (5) for a $\lambda = 800$ nm, $P = 6.7$ PW, $\tau = 30$ fs pulse focused by an $f_\# = 20$ parabola and incident on an $f_\# = 20$ focusing plasma lens of thickness $\Delta = 300$ μm placed 5 mm before the parabola focus. Note that Fig. (2) does not include any nonlinear effects. Asynchronous focusing drops the peak intensity by nearly a factor of two.

**B. Nonlinear modifications**

To illustrate how nonlinearities can modify the focusing properties of the plasma lens, we examine limits of the quasi-static equations commonly used in the modeling of laser wakefield accelerators [20-24]. The quasi-static approximation exploits the disparate time scales of laser pulse and plasma evolution. In particular, the potentials are separated into fast components, varying on the time scale of the laser period, and slow components, varying on the time scale of the pulse duration or plasma period. Additionally, the laser pulse profile is assumed static during the transit time of an electron through the pulse. As we will see, the typical spot size incident on a plasma lens is large, $k_{p0} w_l \gg 1$, where $k_{p0} = (e^2 n_0 / \varepsilon_0 m_e c^2)^{1/2}$. In this limit, the quasi-static equations, correct to order $(k_{p0} w_l)^{-3}$, reduce to the following:

$$\left(\nabla_\perp^2 + \frac{\partial^2}{\partial \xi^2}\right)\phi = k_p^2 (\gamma \rho - \rho^0) \quad (6a)$$

$$\mathbf{u}_\perp = -\frac{1}{k_p^2 \rho} \frac{\partial}{\partial \xi} \nabla_\perp \phi \quad (6b)$$

$$\gamma = \frac{\gamma_\perp^2 + u_\perp^2 + (1+\psi)^2}{2(1+\psi)} \quad (6c)$$

$$\gamma - u_z - \psi = 1 \quad (6d)$$

$$\rho = \frac{\rho^0 + k_p^{-2} \nabla_\perp^2 \psi}{1+\psi} \quad (6e)$$

where $u_z = (\gamma^2 - \gamma_\perp^2 - u_\perp^2)^{1/2}$, $\gamma_\perp = (1 + \tfrac{1}{2}\hat{a}_l^2)^{1/2}$, $\psi = \phi - a_z$, all quantities are understood to vary slowly in time, and potentials are normalized to $m_e c^2/e$.

In the adiabatic (steady-state) limit, $\omega_{p0}\tau \gg 2\pi$, the slow momenta and vector potential are nearly zero, providing $\psi \sim \phi \sim \gamma - 1$ and the fully nonlinear result $\rho = \gamma_\perp^{-1}(\rho^0 + k_p^{-2}\nabla_\perp^2 \gamma_\perp)$. For a thin plasma lens (see Appendix A), the laser pulse profile does not evolve as it passes through the lens, and acquires a phase

$$\Phi(r,\xi) = -i\frac{\Delta}{2k_0\gamma_\perp}\left(k_{pe}^2 + \nabla_\perp^2 \gamma_\perp\right), \quad (7)$$

where $k_{pe}^2 = k_{p0}^2 \pm 4r^2/w_m^4$. The last term on the right hand side of Eq. (7) represents the transverse ponderomotive expulsion of electrons from the laser pulse path. The $\gamma_\perp^{-1}$ coefficient includes the nonlinearity responsible for relativistic self-focusing in the weakly nonlinear limit described above. The intensity dependence of the first and last terms in Eq. (7) imparts a non-quadratic spatial phase to the pulse, giving rise to spherical and higher order optical aberrations.

Recalling that $k_{p0}w_l \gg 1$, we can recover the 1D quasi-static equations by taking $\nabla_\perp \to 0$. In this limit, the slow quantities $\rho$ and $\phi$ only depend parametrically on the transverse coordinate. Following Reference [22], we further assume $|\phi| \ll 1$ to find

$$\rho = 1 - \frac{k_{pe}}{2}\int_0^\xi |\hat{\mathbf{a}}(r,\xi,z)|^2 \sin[k_{pe}(\xi - \xi')]d\xi'. \quad (8)$$

The two terms in Eq. (8) represent the unmodified plasmas density and the excitation of plasma waves by the longitudinal ponderomotive force. Using Eq. (8), one can show that a short, quasi-nonlinear pulse, $\omega_{p0}\tau \ll 2\pi$ and $(k_{p0}c\tau a_l/4)^2 \ll 1$, of the form $\hat{a}_l^2 = \bar{a}_l^2(r)\sin^2(\pi\xi/2c\tau)$ where $0 \leq \xi \leq 2\tau$, acquires a phase

$$\Phi = -i\frac{\Delta}{2k_0}k_{pe}^2\left\{1-\frac{1}{16}k_{pe}^2\bar{a}_l^2\left[\xi^2-\left(\frac{2c\tau}{\pi}\right)^2\sin^2\left(\frac{\pi\xi}{2c\tau}\right)\right]\right\}. \quad (9)$$

upon traversing a thin plasma lens. In contrast to Eq. (7), Eq. (9) demonstrates that each ξ-slice of the pulse acquires a nonlinear phase that does not simply track the ξ-dependence of the pulse intensity. This results from the plasma, owing to its finite inertia, being unable to respond instantaneously to the ponderomotive force applied by the pulse. For both the adiabatic and impulsive limits, the ξ-dependence of the phase results in each slice of the pulse focusing at a different axial position, extending the effective confocal region.

A pulse of intermediate duration $\omega_{p0}\tau > 2\pi$ can undergo the stimulated Raman forward scattering (SRFS) instability [24-27]. Equation (8) shows that the pulse acquires a temporal phase modulation from the plasma wave it excites. For large electron densities or pulse amplitudes, this phase modulation quickly develops into an intensity modulation that reinforces the plasma wave excitation. Through this feedback the laser pulse eventually breaks up into several pulses with an approximate duration of the plasma period. In the small amplitude limit, $a_l^2 \ll 1$, the exponentiation, $\Gamma$, of the SRFS instability is given by $\Gamma \sim (2c\tau a_l^2 k_{p0}^3 \Delta / k_0)^{1/2}$ [24]. Unlike the steady state and impulsive limits, when $\Gamma \gg 1$, the SRFS regime can result in a pulse with both phase aberrations and amplitude distortion as it exits the lens.

For ultrashort MPW pulses none of the limits discussed can be considered strictly valid. In the focusing plasma lens examples presented below, $\omega_{p0}\tau < 2\pi$ resembling the impulsive limit, but because of the high intensities incident on the plasma lens $(k_{p0}c\tau a_l / 4)^2 > 1$. At these high intensities, the slow axial vector potential, neglected in arriving at Eq. (8), contributes significantly to the electron current. We can, however, obtain a rough estimate for when the nonlinear currents disrupt plasma lens focusing. Taylor expanding Eq. (8) to second order in radius, we find that plasma lens focusing is completely negated when $a_l^2 = 8\eta/\omega_{p0}\tau$, where $\eta$ is a pulse-shape dependent parameter of order one (for the case above $\eta = 1 - 4/\pi^2$). This expression provides a condition for linear focusing, $a_l^2 \ll 8/\omega_{p0}\tau$.

The performance degradation of the defocusing plasma lenses is somewhat harder to quantify. Defocusing plasma lenses have an on-axis density maximum, such that even for ultrashort pulses $\omega_{p0}\tau$ can exceed $2\pi$. This resembles the SRFS limit, but because of the high intensities incident on the lens the instability quickly evolves past the linear regime. Based on the simulations in the next section, an SRFS exponentiation of ten provides a rough threshold, $a_l^2 = 50k_0/(c\tau\Delta k_p^3)$, but ultimately this depends on the seeding of the instability. If the on-axis density is lowered, such that $\omega_{p0}\tau < 2\pi$ the condition described for the focusing plasma lens can be used.

**IV. Simulation Results**

As discussed above, characterizing the focal properties of plasma lenses for ultrashort MPW pulses requires fully nonlinear, time-dependent models. Three-dimensional, particle-in-cell (PIC) simulations offer one option [28,29]. The computation times, however, can become unrealistic when optimizing the plasma lens for a particular application: nonlinear plasma lens focusing involves a number of pulse and plasma parameters, and suffers a disparity of length scales inherent in strong focusing geometries. Parameterization of the plasma lens is further discussed in Appendix B.

Simulations of strong focusing geometries must have a transverse domain large enough to enclose the lens diameter and grid spacing sufficient to resolve the focal spot. Defining $N_f$ as the number of cells required to resolve the focal spot, and using the scaling for $D_l$ above, we find that the number of computational cells in one transverse dimension alone scales as $N_\perp \sim 3N_f(P/P_*)^{1/2}(a_l f_\#)^{-1}$ where $f_\#$ is the effective f-number of the plasma lens. As an example, we take P = 10 PW, $N_f$ = 15, $f_\# = 20$, and $a_l$ = 2 to find $N_\perp \sim 1200$. Even for this marginally resolved case the number of transverse cells is sizable.

For computationally efficiency, we simulate the parabola-plasma lens focusing using ponderomotive guiding center (PGC) simulations [30]. The electrons' trajectories evolve according to laser cycle-averaged equations of motion, which include the ponderomotive force and electromagnetic forces associated with the 'slow' wake potentials. The laser envelope is evolved according to Eq. (1). Barring any asymmetry in the initial conditions, the cycle-averaged dynamics are axially symmetric, permitting 2D cylindrical simulation domains. This and the relaxation of the time step allow parabola-plasma lens simulations to be completed in a few hours on a handful of high performance computing nodes.

We examine two focal geometries starting with an $f_\# = 20$ parabola: a focusing plasma lens for an effective $f_\# = 10$ system, and a defocusing plasma lens for an effective $f_\# = 30$ system. The $f_\# = 20$ parabola and laser pulse parameters are motivated by a beam path of the planned MPW ELI-NP Ti-Sapphire laser system [15]. Specifically, we consider a 10 m focal length, 0.5 m diameter parabola focusing a spatio-temporal Gaussian pulse with $\lambda = 800$ nm and $\tau = 30$ fs. In all cases, the laser pulse is initialized with the appropriate phase front curvature and amplitude acquired by the aforementioned ELI parabola. In the absence of a plasma lens, the focus occurs at $z = 0$ and the focal spot is $w_f = 19.5$ μm. All plasma lenses considered have a thickness of $\Delta = 300$ μm.

**A. Focusing Plasma Lens**

The focusing plasma lenses were $f_\# = 20$ making the combined parabola-plasma lens system an $f_\# = 10$. The lenses had an initial radial profile $n_e = n_{\min} + r^2 / \pi r_e w_m^4$ for $r < k_{pr} w_m^2 / 2$ and $n_e = n_{\max}$ for $r \geq k_{pr} w_m^2 / 2$ where $k_{pr} = [e^2(n_{\max} - n_{\min}) / \varepsilon_0 m_e c^2]^{1/2}$. The on-axis and peak densities were fixed at $n_{\min} = 1 \times 10^{18}$ cm$^{-3}$ and $n_{\max} = 1.2 \times 10^{20}$ cm$^{-3}$ respectively. Simulations, not presented, indicated that the results are insensitive to a lowering of the on-axis density. Figure (3a) displays the transverse density profile.

Figure 4 shows the peak intensity as a function of pulse power for two plasma lens locations: (a) $z_l = -0.5$ cm and (b) $z_l = -1.0$ cm. We note that $w_m$ was changed with the lens location to maintain the f-number. In particular, $w_m = 17.6$ $\mu$m for $z_l = -0.5$ cm and $w_m = 21$ $\mu$m for $z_l = -1$ cm. The black solid line (top), blue solid line (middle), and red solid line (bottom) are the peak intensity predictions for ideal plasma lens focusing, asynchronous plasma lens focusing, and parabola-only focusing respectively. The green dashed line and circles demarcate simulation results. The spot incident on the $z_l = -0.5$ cm lens was $w_l = 250$ $\mu$m, while the incident $a_l$ varied with the power from 0.36 to 5.2. For $z_l = -1.0$ cm, $w_l = 250$ $\mu$m with $a_l$ ranging from 0.18 to 2.6.

For low powers, the drop in peak intensity resulting from asynchronous focusing can be observed by comparing the black (top) and blue (middle) curves. Below $P = 700$ TW the predictions of Eq. (4) are in excellent agreement with the simulations. Consistent with our condition, $(c\tau / w_l) f_\# < 1$, the asynchronous focusing intensity reduction becomes more severe the further the plasma lens is placed from the parabola's focus: $w_l = 125$ $\mu$m for $z_l = -0.5$ cm and $w_l = 250$ $\mu$m for $z_l = -1.0$ cm. At higher powers, the pulse drives nonlinear currents sufficient to distort the phase front curvature applied by the plasma lens. A power of 2 PW, for instance, corresponds to $a_l = 2$ and $a_l = 1$ for $z_l = -0.5$ cm and $z_l = -1.0$ cm respectively. When the power surpasses $P = 3$ PW, the plasma lens reduces the peak power beyond that of the parabola alone.

Figure 5 displays the transverse Laplacian of the phase applied by the $z_l = -0.5$ cm plasma lens at the $\xi$-slice of peak intensity for three pulse powers: $P = 0.14$ PW blue (nearly horizontal), $P = 1.4$ PW green (small fluctuation), and $P = 14.0$ PW red (large fluctuation). Specifically the plot shows $-(\Delta k_p^2 / 2k_0) \nabla_\perp^2 \rho(\mathbf{r}, \xi_{max})$, where $[\partial_\xi I(\mathbf{r}, \xi)]_{\xi=\xi_{max}} = 0$. A constant, horizontal line would indicate an ideal thin lens transverse phase. The transverse intensity profile has been plotted for reference as the dashed black line.

The drop in the vertical intercept with increasing power indicates a reduction in focusing. In particular, the effective focal length has dropped by a factor of 1.3 for $P = 1.4$ PW and 3.7 for $P = 14$ PW, corresponding to peak intensity reductions of 1.7 and 14 respectively. This is larger than the peak intensity drop shown in Fig. (4a). At higher powers, ξ-slices starting earlier in the pulse experience less nonlinearity and reach higher peak intensities. In addition to the focusing reduction, the parabolic and higher order radial dependence observable for $P = 1.4$ PW and $P = 14$ PW indicate that the nonlinear currents contribute to spherical and higher order aberration.

**B. Defocusing Plasma Lens**

The defocusing plasma lens parameters were chosen to make the combined parabola-plasma lens system an $f_{\#} = 30$. The lenses had a radial profile $n_e = n_{max} - r^2 / \pi r_e w_m^4$ for $r < k_{pr} w_m^2 / 2$ and $n_e = n_{min}$ for $r \geq k_{pr} w_m^2 / 2$. The minimum density was fixed at $n_{min} = 1 \times 10^{18}$ cm$^{-3}$. Several peak densities were simulated to examine its affect on the defocusing. Figure (3b) displays the transverse density profile for $n_{max} = 4 \times 10^{19}$ cm$^{-3}$.

Figure (6a) shows the peak intensity as a function of pulse power for a defocusing plasma lens with $n_{max} = 4 \times 10^{19}$ cm$^{-3}$ located at $z_l = -1.0$ cm. The location was chosen to reduce the effect of nonlinearities and to maintain a reasonably sized plasma lens. As with the focusing lens, the incident spot size was $w_l = 250$ μm with $a_l$ ranging from 0.18 to 2.6. The red solid line (top) shows the peak intensity for parabola-only focusing. The black solid and blue solid lines (bottom, essentially overlayed) are the peak intensity predictions for ideal and asynchronous plasma lens focusing respectively. The green dashed line and circles demarcate simulation results.

The simulations points fall nearly on top of the ideal plasma lens theory predictions. Unlike the focusing plasma lens, the defocusing plasma lens shows no reduction in intensity from asynchronous focusing. This is consistent with Eq. (4). As discussed above, for a focusing plasma lens, the diffractive and thin lens phases add essentially doubling the chromatic aberration. For the defocusing plasma lens these phases are of opposite sign and act to reduce the chromatic aberration.

Even at high powers Fig. (6a) shows the defocusing plasma lens following the linear theory. In Fig. (6b), however, we see that the position of peak intensity moves closer to the plasma lens as the power increases—the plasma lens is lowering the intensity, but not producing the desired refraction. This reduction in refraction can be observed in the phase applied by the lens. Figure (7a) displays the transverse Laplacian of the phase acquired near the exit of the plasma lens at the ξ-slice of peak intensity for $P = 3.1$ PW in red (fluctuations) and an ideal defocusing plasma lens in green (horizontal line). As with Fig. 5, the plot shows specifically $-(\Delta k_p^2 / 2k_0)\nabla_\perp^2 \rho(\mathbf{r}, \xi_{max})$, where $[\partial_\xi I(\mathbf{r},\xi)]_{\xi=\xi_{max}} = 0$. The transverse intensity profile has been plotted for reference as the dashed black line. Near the axis, the curvature is almost zero, negating the plasma lens refraction.

While Fig. (7a) illustrates the flattening of the transverse phase at one axial location, pulse evolution in the defocusing lens leads to axial variations in the phase. This is in contrast to the focusing lens where the thin lens approximation could be considered valid. In particular, the pulse quickly enters the nonlinear regime of stimulated Raman forward scattering. Figure (7b) shows the on-axis intensity profile of the pulse at the entrance and exit of the lens in black (smooth) and red (fluctuating) respectively. The intensity peaks spaced at the plasma period indicative of Raman forward scattering are evident. Both phase aberrations and intensity distortion inhibit the refraction and focal quality of the defocusing plasma lens.

Based on the SRFS exponentiation presented in the last section, we can attempt to "correct" the intensity fluctuations by dropping the peak density of the lens. With the plasma lens curvature fixed by the desired focal geometry, a drop in peak density requires a decrease in the lens diameter: $D_l = k_{pr} w_m^2$ where $k_{pr} = [e^2(n_{max} - n_{min})/\varepsilon_0 m_e c^2]^{1/2}$. The diameter cannot be made too small, however, or the pulse power will be clipped. In particular, the power refracted by the plasma lens is given by $P_r = P[1 - \exp(-k_{pr}^2 w_m^4 / 2w_l^2)]$. We consider a "Raman-corrected" defocusing plasma lens with a peak density $n_{max} = 1 \times 10^{19}$ cm$^{-3}$ that refracts 80% of the pulse power. Figure (8a) shows the transverse density profile of this lens as the dashed blue curve, and the original, $n_{max} = 4 \times 10^{19}$ cm$^{-3}$, lens as the solid red line. Note that both lenses have the same curvature, but different diameters.

The absence of SRFS intensity modulations is clear in Fig. (7b), which displays the on-axis pulse profile near the exit of the Raman-corrected lens as the blue dashed line. Figure (7a) shows that lowering the density has also reduced the focusing and higher order aberrations. Even with this reduction, other nonlinearities, such as discussed in the previous section, result in the peak intensity occurring too early and with a wide effective confocal region. Figure (8b) displays the on-axis fluence of the $P = 3.1$ PW pulse for the original defocusing lens in red (solid) and the Raman-corrected lens in blue (dashed). A scaled version of the $P = 6.7$ TW fluence resulting from the original lens is also plotted as an example of a near-linear plasma lens defocusing.

## V. Summary and conclusions

We have examined the time-dependent and nonlinear effects that modify plasma lens focusing of ultrashort pulse MPW pulses. We demonstrated that focusing plasma lenses suffer from both chromatic and nonlinear phase aberrations. For ultrashort pulses, chromatic aberration results in asynchronous focusing manifesting as a boomerang like pulse profile and a corresponding drop in peak intensity. The nonlinear phase aberrations, predominately focusing and spherical, also reduced the peak intensity. While the plasma lens provided enhanced focusing up to ~1 PW, the performance degraded through the MPW regime, reaching a focusing penalty at ~10 PW. We note that the asynchronous focusing will be less acute for longer pulses, such as those produced by Nd:glass lasers, but these longer pulses are also more susceptible to plasma instabilities. Detailed examination of this issue is left for future work.

Defocusing plasma lenses, while robust to chromatic aberration, suffer from both phase aberration and amplitude distortion resulting from nonlinear stimulated Raman forward scattering (SRFS). The defocusing lenses produced the desired focal intensity and approximate focal point up to ~1 PW, but degraded rapidly in the MPW regime. Reducing the lens diameter was shown to reduce the amplitude distortion associated with SRFS at the cost of clipping pulse power, but the remaining phase aberrations remained sizable.

In contrast to the thin slab, parabolic profile lenses considered here, one can imagine spatially structuring plasma lenses either radially, to correct for spherical aberration, or longitudinally to correct for chromatic aberration, a parabolic plasma mirror for instance. Additionally, one may consider more exotic solutions such as a radial spatial chirp for compensating chromatic aberration or a plasma-based Fresnel lens to reduce the role of density-dependent aberrations. Our continuing research will examine these alternatives.

**Acknowledgements**

The authors would like to thank Y.-H. Chen and A. Ting for fruitful discussions. This work was supported by the Naval Research Laboratory 6.1 Base Program.

**Appendix A: Conditions on Thin Lens Approximation**

The thin lens approximation simplifies calculations of the plasma lens focusing properties, but comes with caveats. These amount to three conditions: the pulse cannot focus within the plasma lens itself, the pulse intensity cannot change appreciably within the plasma lens, and the pulse cannot undergo substantial longitudinal evolution within the lens. The first two, associated with the refraction and diffraction of the pulse, imply $\Delta < k_0 w_m^2 / 4$ and $\Delta \ll z_{pl}$ respectively. Longitudinal evolution can result from spectral shifting and depletion of the pulse within the plasma, implying $\Delta < a_l (k_0 / k_p)^2 k_p^{-1}$ [31]. For pulses long compared to the plasma period, the stimulated Raman forward scattering instability can also contribute to longitudinal evolution. In this regime, a thin lens requires $\Delta < (k_0 / 2c\tau a_l^2 k_p^3)$ [24]. The first three conditions are not particularly restrictive. With typical Ti-Sapphire parameters, $\lambda = 800$ nm, $w_m = 14$ $\mu$m, $n_e = 1 \times 10^{18}$ cm$^{-3}$, and $a_l = 1$, we find $\Delta < 0.5$ mm for the refractive condition and $\Delta < 1$ cm for pulse depletion. As discussed in the main text, the condition on Raman scattering can be more difficult to satisfy for defocusing lenses.

**Appendix B: Parameterizing of the Plasma Lens**

In the time-independent linear regime, only two parameters determine the evolution of a Gaussian beam focused by a thin plasma lens: $z_l / Z_R$ and $(w_f / w_m)^4 (\Delta / Z_R)$ where $z_l$ is the distance of the plasma lens from the parabolic mirror focus and $Z_R = k_0 w_f^2 / 2$ is the Rayleigh length [5]. These specify the initial condition of the Gaussian beam incident on the plasma lens and the phase front curvature applied by the plasma lens respectively. Weakly nonlinear pulses can undergo Kerr-like self-focusing due to the effective relativistic mass increase of electrons oscillating in the laser field [32]. This introduces two additional parameters. The first, $(P / P_{cr})(\Delta / Z_R)$, where $P_{cr} = 17(\omega / \omega_{p0})^2$ GW is the self-focusing critical power and $\omega_{p0} = (e^2 n_0 / \varepsilon_0 m_e)^{1/2}$, quantifies the nonlinear phase curvature imparted to the pulse as it traverses the lens. The second, $(P / P_{cr})(w_f / k_{p0} w_m^2)(\Delta / Z_R)$ where $k_{p0} = \omega_{p0} / c$, accounts for the variation in the nonlinear phase curvature with density [33]. The last three parameters allow one to quickly assess the relative importance of each contribution to the phase front curvature. This Kerr-like regime, however, omits time-dependence and the ponderomotive force, restricting it to a narrow range of parameter space for which $\omega_{p0}\tau \gg 1$ and $k_{p0}w_l \gg 1$.

For the highly nonlinear ultrashort pulses characteristic of MPW lasers, the ponderomotive force dynamically deforms the electron density, and, in some cases, cavitates the electrons altogether [34]. While $P/P_{cr}$ can serve as a gauge of nonlinearity in this regime, its absence from the governing equations diminishes the utility of the weakly nonlinear parameters defined above. The result is that six parameters are required to determine the focusing properties of a thin plasma lens for MPW systems: $z_l/Z_R$, $\Delta/Z_R$, $w_f/w_m$, $k_{p0}w_f$, $a_l$, and $c\tau/w_f$.

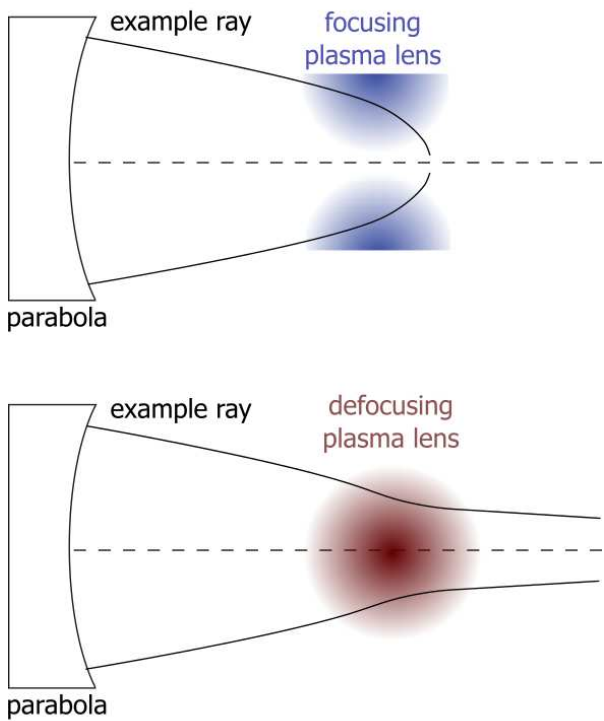

Figure 1. Schematics of combined solid-state parabolic mirror and plasma lens focusing systems. The plasma density is minimum on axis for the focusing plasma lens, and maximum on axis for the defocusing plasma lens.

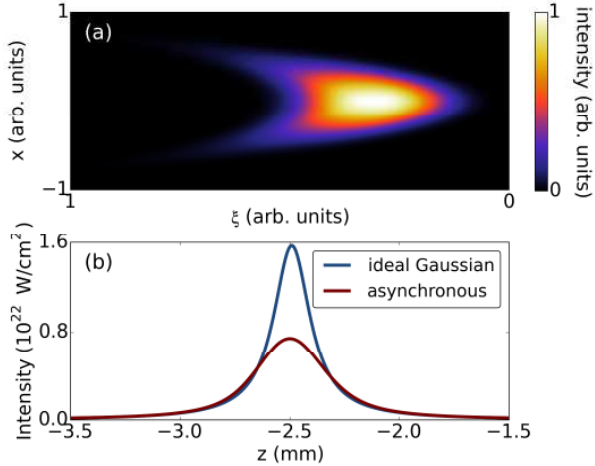

figure 2. (a) An intensity profile in the x-ξ plane characteristic of asynchronous focusing—a result of plasma lens dispersion. (b) On-axis intensity for an ideal Gaussian beam (blue) and an ultrashort pulse (red) for a $\lambda = 800$ nm, $P = 6.7$ PW, $\tau = 30$ fs pulse focused by an $f_\# = 20$ parabola and incident on an $f_\# = 20$ plasma lens of thickness $\Delta = 300$ μm placed 5 mm before the parabola focus. The reduction in intensity due to asynchronous focusing is clear.

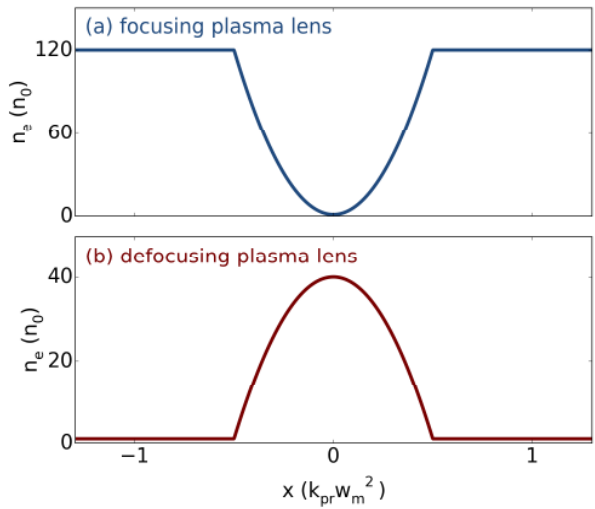

Figure 3. (a) and (b) Transverse density profiles used for the focusing and defocusing plasma lens simulations respectively.

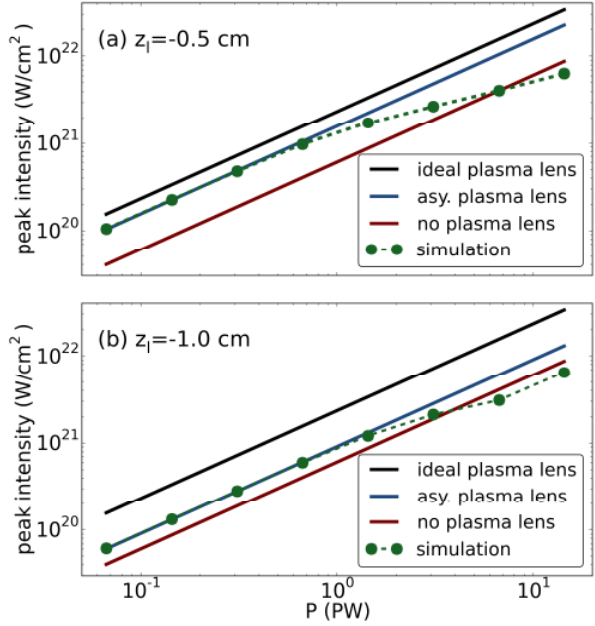

Figure 4. Peak intensity as a function of pulse power for two focusing plasma lens locations: (a) $z_l = -0.5$ cm and (b) $z_l = -1.0$ cm. The black solid line (top), blue solid line (middle), and red solid line (bottom) are the peak intensity predictions for ideal plasma lens focusing, asynchronous plasma lens focusing, and parabola-only focusing respectively. The green dashed line and circles demarcate simulation results.

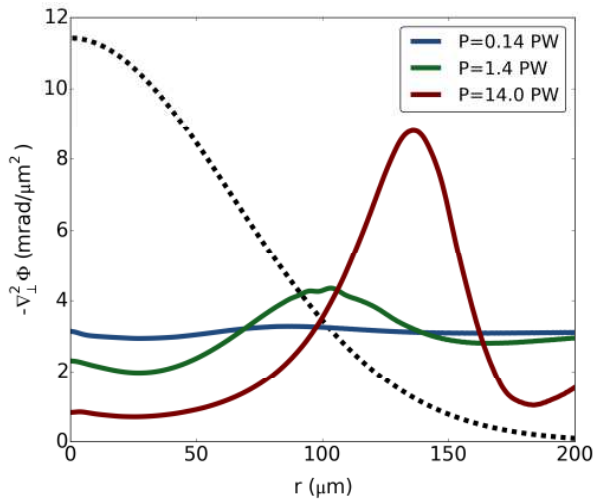

Figure 5. Transverse Laplacian of the $z_l = -0.5$ cm plasma lens phase at the ξ-slice of peak intensity for three pulse powers: $P = 0.14$ PW blue (nearly horizontal line), $P = 1.4$ PW green (small fluctuations), and $P = 14.0$ PW red (large fluctuations). A constant, horizontal line would indicate an ideal thin lens transverse phase. For reference, the transverse intensity profile in arbitrary units is displayed as the dashed black line.

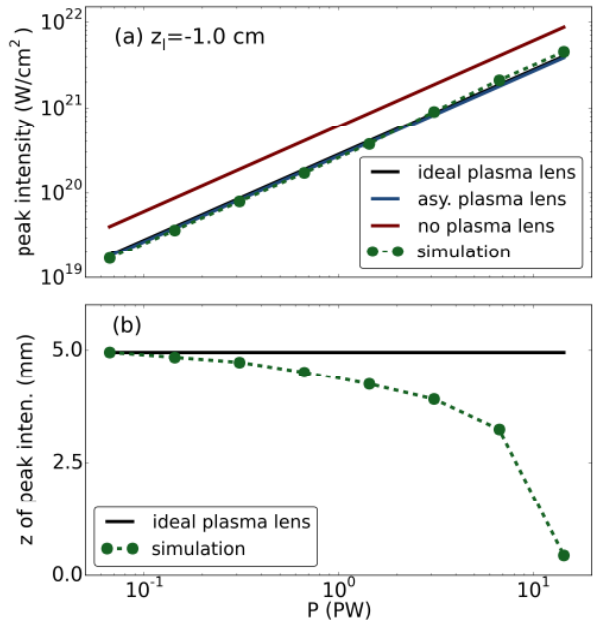

Figure 6. (a) Peak intensity as a function of pulse power for a defocusing plasma lens at $z_l = -1.0$ cm. The red solid line (top) shows the peak intensity for parabola-only focusing. The black solid and blue solid lines (bottom, essentially overlayed) are the peak intensity predictions for ideal and asynchronous plasma lens focusing respectively. The green dashed line and circles demarcate simulation results. (b) Axial location of the peak intensity as a function of power. The focal position of the parabola is at zero.

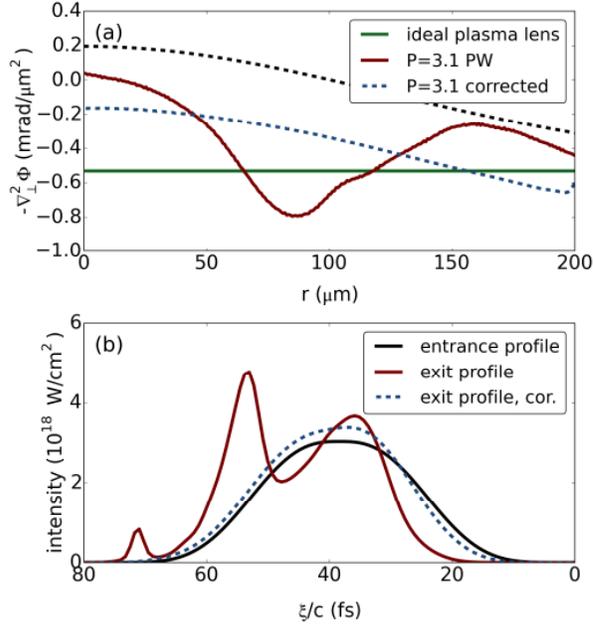

Figure 7. (a) Transverse Laplacian of the defocusing plasma lens phase at the $\xi$-slice of peak intensity for an ideal plasma lens, green solid line (horizontal), the $n_{max} = 4\times 10^{19}$ cm$^{-3}$ lens, red solid line (large fluctuations), and the Raman-corrected $n_{max} = 1\times 10^{19}$ cm$^{-3}$ lens, blue dashed line. For reference, the transverse intensity profile in arbitrary units is displayed as the dashed black line. (b) On-axis intensity profile at the entrance of plasma lens, black solid line (smooth), near the exit of the plasma lens for the $n_{max} = 4\times 10^{19}$ cm$^{-3}$ lens, solid red line (fluctuations), and the Raman-corrected $n_{max} = 1\times 10^{19}$ cm$^{-3}$ lens, blue dashed line.

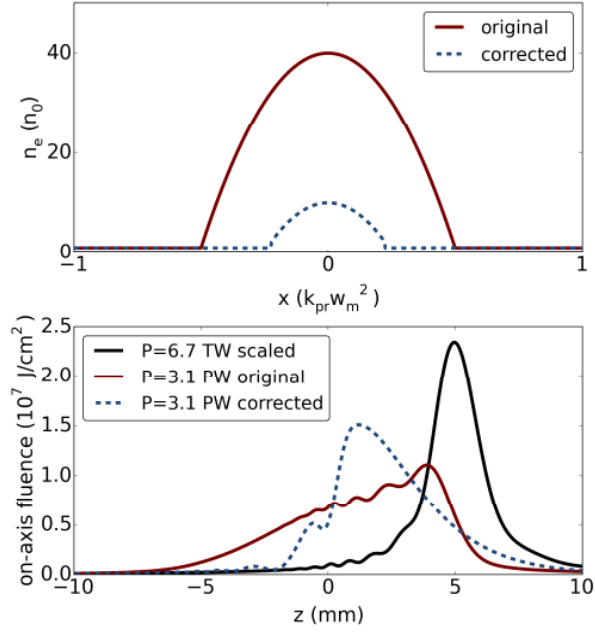

Figure 8. (a) Transverse density profiles for the $n_{max} = 4\times10^{19}$ cm$^{-3}$ defocusing plasma lens, solid red, and the Raman-corrected $n_{max} = 1\times10^{19}$ cm$^{-3}$ defocusing plasma lens, dashed blue. (b) On-axis fluence for a $P = 3.1$ PW pulse refracted by the $n_{max} = 4\times10^{19}$ cm$^{-3}$ and $n_{max} = 1\times10^{19}$ cm$^{-3}$ plasma lenses, solid red and dashed blue respectively. The scaled on-axis fluence of a $P = 6.7$ TW pulse refracted by the $n_{max} = 4\times10^{19}$ cm$^{-3}$ lens is shown as an example of near-linear plasma lens defocusing.